\font\bfmath=cmmib10
\def\bfvarepsilon{\hbox{\bfmath\char'042}}
\def\mK{{\rm \mu K}}

\def\expec#1{\langle#1\rangle}

\def\etal{{\frenchspacing\it et al.}}
\def\ie{{\frenchspacing\it i.e.}}
\def\eg{{\frenchspacing\it e.g.}}
\def\etc{{\frenchspacing\it etc.}}

\def\pp{\noindent\parshape 2 0truecm 13.6truecm 1truecm 12.6truecm}
\def\rf#1;#2;#3;#4 {\par\pp#1, {\it #2}, {\bf #3}, #4. \par}
\def\rg#1;#2;#3;#4;#5 {\par\pp#1, {\it #2}, {\bf #3}, #4 (``#5"). \par}
\def\rn{\pp}

\def\beq#1{\begin{equation}\label{#1}}
\def\eeq{\end{equation}}
\def\beqa#1{\begin{eqnarray}\label{#1}}
\def\eeqa{\end{eqnarray}}
\def\eq#1{equation~(\ref{#1})}

\def\bfig{\begin{figure}[h] \centerline{\hbox{}}\vfill}
\def\efig{\end{figure}\vfill\newpage}

\def\fheight{12cm}
\def\fwidth{17cm}
\def\fig#1{Figure~\ref{#1}}

\def\spose#1{\hbox to 0pt{#1\hss}}
\def\simlt{\mathrel{\spose{\lower 3pt\hbox{$\mathchar"218$}}
     \raise 2.0pt\hbox{$\mathchar"13C$}}}
\def\simgt{\mathrel{\spose{\lower 3pt\hbox{$\mathchar"218$}}
     \raise 2.0pt\hbox{$\mathchar"13E$}}}
\def\simpropto{\mathrel{\spose{\lower 3pt\hbox{$\mathchar"218$}}
     \raise 2.0pt\hbox{$\propto$}}}

\def\addr#1{{\small\it #1}}
\def\auth#1{{#1}}

\def\vx{{\bf x}}
\def\vy{{\bf y}}
\def\vz{{\bf z}}
\def\vxp{\tilde{\bf x}}
\def\Zp{\tilde Z}
\def\Mp{\tilde M}
\def\va{{\bf a}}

\def\vn{{\bfvarepsilon}}
\def\vq{{\bf b}}
\def\nh{\widehat{\bf n}}
\def\N{N}
\def\Cn{C^{(n)}}
\def\lstar{l_0}
\def\L{L}
\def\Q{Q}
\def\Qrmsps{Q_{rms-PS}}

\def\expec#1{\langle#1\rangle}

\documentstyle[11pt,epsf]{article}
\begin{document}


\begin{titlepage}   

\noindent
April 22, 1995
\hfill MPI-PhT/94-89
\begin{center}

\vskip0.9truecm
{\bf
A BRUTE FORCE ANALYSIS OF THE COBE DMR DATA\footnote{
Published in {\it ApJ}, {\bf 455}, {\it 1-6}, December 10, 1995.
Submitted December 5, 1994.\\
Available from\\
{\it h t t p://www.sns.ias.edu/$\tilde{~}$max/brute.html} 
(faster from the US) and from\\
{\it h t t p://www.mpa-garching.mpg.de/$\tilde{~}$max/brute.html} 
(faster from Europe).\\
}
}

\vskip 0.5truecm
  \auth{Max Tegmark}
  \smallskip

  \addr{Max-Planck-Institut f\"ur Physik, F\"ohringer Ring 6,}
  \addr{D-80805 M\"unchen;}

 \addr{e-mail: max@ias.mpg.de}

  \smallskip
  \vskip 0.2truecm

  \auth{Emory F. Bunn}

  \smallskip
  \addr{Department of Physics, University of California,}
  \addr{Berkeley, CA 94720;}

  \addr{e-mail: bunn@leporello.berkeley.edu}
  \smallskip

\end{center}

\abstract{
More than a dozen papers analyzing the COBE data have now appeared.
We review the different techniques
and compare them to a ``brute force"
likelihood analysis
where we invert the full $4038\times 4038$ Galaxy-cut pixel
covariance matrix.
This method is optimal in the sense of producing
minimal error bars, and is a useful reference point for comparing
other analysis techniques.
Our maximum-likelihood estimate of the spectral index and normalization
are $n \approx 1.15 (0.95)$ and $\Q \approx 18.2 (21.3)\, \mK$
including (excluding) the
quadrupole. Marginalizing over the normalization $C_9$, 
we obtain
$n\approx 1.10\pm 0.29$ ($n\approx 0.90\pm 0.32$).
When we compare these results with those of 
the various techniques that involve
a linear ``compression'' of the data, we find that the latter 
are all consistent with the brute-force analysis and have error bars
that are nearly as small as the minimal error bars.  We therefore conclude
that the data compressions involved in these techniques do indeed retain
most of the useful cosmological information.
}

\end{titlepage}


\section{Introduction}

Since the first cosmic microwave background (CMB) 
anisotropies were detected by the 
COBE DMR experiment 
(Smoot {\etal} 1992), a plethora of different analysis techniques have
been published.
The aim of this paper is to compare them to a 
computationally cumbersome 
but statistically optimal method, 
to see how good they are.
This is quite timely given the rate at which our
data about the CMB sky is accumulating, since we wish to 
employ analysis techniques which are fast 
when faced with large data volumes and at the same time 
give error bars that are near the 
theoretical minimum.  

All published COBE analysis techniques have involved two steps: 

\begin{enumerate}

\item
The full Galaxy-cut data set, consisting of say $N=4038$ numbers, 
is by
some clever form of ``data compression"
reduced to a smaller data set with 
$N'<N$ numbers. 

\item 
Cosmological parameters are constrained by analyzing this reduced 
data set, usually by 
computing the likelihood function $L(n,\Q)$.

\end{enumerate}
The reason for doing the data compression is to speed up the 
calculations in 2, which would otherwise involve 
repeated inversions of $4038\times 4038$ matrices in 
the case of the COBE DMR experiment.
The idea is that if the compression method cleverly takes
the physics into account, it mostly throws away noise and keeps
the bulk of the cosmological information. 

Bond (1994, hereafter B95),
G\'orski {\etal} (1994a, 1994b, hereafter G94), and Bunn \& Sugiyama (1995,
hereafter BS95) 
all use {\it linear} data compression, where the reduced data is 
simply the original data vector multiplied by some matrix.
Their reduced data sets contain 928, 957 and 400
numbers, respectively. In G94, the 
row vectors of the compression matrix are chosen to be those 
corresponding to the multipoles $l=2-30$, weighted 
with account to pixel noise and orthogonalized.
In BS95, the 
row vectors are instead those that arise from a certain 
eigenvalue problem.  [This technique is described further in Bunn 
\etal (1995) and White \& Bunn (1995).]
This data compression technique is often referred
to as ``expansion in signal-to-noise eigenfunctions"
and is known in signal processing as the Karhunen-Lo\`eve expansion
(Karhunen 1947).
Given a prescribed ``compression factor" $N/N'$, it
can be shown to be
optimal in a certain sense (BS95).
This method was independently introduced into cosmology in B95,
where, after first reducing the data by a factor 4 by
smoothing\footnote{The ``pixel level 5" data set was used, where 
each pixel is obtained by averaging four pixels from the 6144 pixel 
data set.},
the effect of compressing further
with various $N'$ is studied in detail.

A second group of methods use {\it quadratic} data compression,
where the reduced data are various quadratic combinations of the
pixel values.
Work in this category includes 
Smoot {\etal} (1992), Bennett {\etal} (1992),
Scaramella \& Vittorio
(1993), Seljak \& Bertschinger (1993), and Bennett {\etal} (1994),
where the reduced data consist of bins of the
observed correlation function, typically around 60. 
Wright {\etal} (1994b) and de Oliveira-Costa \& Smoot (1995) 
use another set of quadratic statistics, 
the first 30 multipole moments of the data.
Estimates of the quadrupole (Gould 1993) and the
total pixel variance $(\Delta T/T)^2$
(Banday {\etal} 1994, Wright {\etal} 1994a)
also fall into the quadratic category.

A third group of methods 
compress the data set by forming quantities that are 
higher-order combinations of the data
than the above-mentioned linear and quadratic ones.
This includes cubic quantities (Hinshaw {\etal} 1994), correlation of extrema
(Kogut {\etal} 1995) or
topological information such as genus and spot
morphology (Smoot {\etal} 1994, Torres 1994)
as the reduced data.

What do we mean by an analysis technique being good?  Apart from the
obvious requirement that it should give fairly unbiased estimates of
the cosmological parameters, we want it to give as small error bars as
possible. This is, of course, but one of several criteria
one could choose to adopt.  In particular, one could aim for a
technique that was maximally robust to systematic errors such as
residual Galactic contamination, or for a technique that
was numerically inexpensive to implement.  One might also decide to prefer
techniques which are as model-independent as possible.  Adopting the
latter criterion might lead one to look unfavorably on the technique
of BS95, as that method involves the choice of a ``fiducial power
spectrum'' in choosing the basis functions.  However, the ``brute
force'' technique described herein does fare well by this criterion,
since it assumes only that the CMB fluctuations are Gaussian.  We have
chosen not to consider the effect of systematic errors or Galactic
contamination in this paper, although the reader should of course 
bear in mind that 
the choice of the data analysis technique used to derive
the cosmological parameters and their uncertainties may not be as
significant as unmodeled systematic effects in the data, such as 
Galactic emission.
We also chose to disregard computational
complexity as a factor when chosing a data analysis method,
in the spirit that the computational work in the data analysis
step will in any case be
negligible compared to the amount of effort already spent on 
collecting the data set. In summary, we focus entirely on 
minimizing the statistical error bars. 

The size of the resulting error bars
clearly depends on the choice of data compression method.  One would
expect that the smallest error bars would result from an analysis in
which no compression at all was performed.  
In the Appendix we
show that all linear compression techniques give likelihood
functions that are on average at least as broad as the likelihood
function arising from an analysis done without compression, and it is
reasonable to expect that the same is true of nonlinear techniques.
For instance, a corollary of the Fisher-Cram\'er-Rao inequality
(Fisher 1935; Kenney \& Keeping 1951 p. 373), well-known to statisticians, implies that if there
exists a best unbiased estimate of the parameters, it will simply be
the maximum likelihood estimate using {\it all} the data.\footnote{
Asymptotically, {\ie}, in
the limit of infinitely large sample size, the conclusion becomes
even stronger: no other method, linear or not, can produce
smaller error bars than a maximum likelihood analysis using all the data. 
Since the COBE data
contain a large number of independent data points --- there are about
100 signal-to-noise eigenmodes with eigenvalue greater than 1 (BS95)
--- we expect the asymptotic limit to be a good approximation.}

We carry out this compression-free
analysis in the present paper, to answer the question 
of how small these optimal error bars are.  
Once this is done, we can easily rank the other methods by checking how close
to optimal their error bars are.
This may be termed ``the brute force approach'', as 
the intuitive simplicity of avoiding data compression
comes at the expense of a significant increase in CPU time.
However, as we discuss below, 
the computations are in fact not as time-consuming as one may think.
Also, since the brute force approach does not require any
time-saving approximations, one can at no extra cost include 
additional elements of realism in the model. 
Thus we investigate the effect of correlated noise 
(Wright {\etal} 1994a; Lineweaver {\etal} 1994)
and the effect of the standard method for dipole removal.
As has been pointed out elsewhere,
the latter tends to bias the estimates 
towards higher values of $n$ if not properly accounted for.  

In techniques based on linear compression, it is easy to account
properly for the effects of monopole and dipole removal, either by
applying the same projection operator to the data covariance matrix
as was applied to the data, or by marginalizing over the unknown modes.
The only way to remove the monopole and dipole bias from estimates
based on nonlinear techniques is to perform Monte Carlo 
simulations.\footnote{
In general, since quantities like $n$ and $\Q$ are not linear in the data,
no technique, linear or otherwise, is guaranteed {\it a priori} to give 
unbiased
parameter estimates; the only way to be sure that a particular technique
is unbiased is to perform Monte Carlo simulations.  Almost
all of the cited
techniques have been tested for bias in this way.}

In the next section, we introduce some simplifying notation.
In section 3, we describe how the pixel correlation 
is affected by removing the monopole, dipole and quadrupole.
In section 4, we present our results, and in the discussion section
we compare them to the previously published techniques.

\section{Notation}

As pointed out by numerous authors, COBE analysis is basically  
linear algebra, and both the equations and their 
interpretation tend to become simplified if we write
the various quantities as vectors and matrices.

Let us write the CMB sky map as the $N$-dimensional vector ${\vx}$, defined by
\beq{xDefEq}
x_i = {\Delta T\over T}(\nh_i),
\eeq
where $\nh_i$ is a unit vector in the direction of the $i^{th}$ COBE pixel.
$\N=6144$ for an all-sky map, and $\N=4038$ after a $20^{\circ}$ Galactic cut.
We write $\vx$ as a sum of three terms,

\beq{xEq}
\vx = Y\va+\vn + Z\vq,
\eeq
which correspond to the contribution from cosmology, 
instrumental noise and ``nuisance" multipoles, respectively,
and will now be described in more detail.

The $N\times\infty$-dimensional spherical harmonic matrix $Y$ 
is defined as 
\beq{Ydef}
Y_{i\lambda} \equiv Y_{lm}(\nh_i),
\eeq
where we have combined $l$ and $m$ into the single index 
$\lambda\equiv l^2+l+m+1 = 1, 2, 3, ...$.
(Throughout this letter, we use real-valued spherical harmonics, which
are obtained from the standard spherical harmonics by replacing $e^{im\phi}$
by $\sqrt 2\sin m\phi$, $1$, $\sqrt 2\cos m\phi$ for $m<0$, 
$m=0$, $m>0$
respectively.)
Making the standard assumption that 
the CMB is Gaussian on COBE scales, $\va$ is an
infinite-dimensional Gaussian random vector
with zero mean and with the diagonal covariance matrix
\beq{Cldef}
\expec{a_{\lambda}a_{\lambda'}} = \delta_{\lambda\lambda'} C_l,
\eeq
the {\it angular power spectrum} $C_l$ being
specified by some cosmological model.
Because of the addition theorem for spherical harmonics,
this leads to the covariance matrix of the cosmological term 
in \eq{xEq} being given 
by 
\beq{CosmoCorrEq}
\expec{Y\va (Y\va)^t}_{ij} = C(\nh_i\cdot\nh_j),
\eeq
where the {\it angular correlation function} is defined as
\beq{CorrEq}
C(\cos\theta) = 
{1\over 4\pi} 
\sum_{l=0}^{\infty} 
(2l+1) C_l P_l(\cos\theta),
\eeq
$P_l$ denoting the $l^{th}$ Legendre polynomial.

The $\N$-dimensional
noise
vector $\vn$ is assumed to be Gaussian with 
$\expec{\vn} = 0$. To an excellent approximation,
its covariance matrix can be
written as 
$\expec{n_i n_j} = \sigma_i\sigma_j \Cn(\nh_i\cdot\nh_j)$,
where $\sigma_i$ is the rms noise of pixel $i$, and
the dimensionless noise correlation function $\Cn$
has been computed by Lineweaver {\etal} (1994). 
To a good approximation, 
$
\expec{n_i n_j} = \sigma_i^2\delta_{ij}
$,
but due to the beam-switching strategy used in the COBE DMR 
experiment, there are some minor corrections to this, primarily
a correlation of the order of $0.5\%$ between pixels separated by 
$60^{\circ}$.

The third term in~\eq{xEq}, the ``nuisance term", 
contains the unknown influence of the multipoles
with $\l \leq \lstar$.
Since we lack accurate a priori knowledge of
both the current CMB temperature and our Galaxy's peculiar
velocity, we need to set $\lstar\geq 1$. 
Some authors are concerned about systematic errors in the
quadrupole and therefore set
$\lstar=2$. The $N\times(\lstar+1)^2$-dimensional matrix $Z$ 
is simply the first $(\lstar+1)^2$ columns of $Y$,
and $\vq$ is a $(\lstar+1)^2$-dimensional constant vector 
whose value we a priori know nothing whatsoever about.
Since the noise $\vn$ and the cosmic signal $\va$ are uncorrelated,
the pixel covariance matrix 
\beq{MdefEq}
M\equiv\expec{\vx\vx^t} - \expec{\vx}\expec{\vx}^t
\eeq
is given by 
\beq{Meq}
M_{ij} = C(\nh_i\cdot\nh_j) + \sigma_i \sigma_j \Cn(\nh_i\cdot\nh_j).
\eeq

\section{On monopole, dipole and quadrupole removal}

\begin{figure}[phbt]
\centerline{\epsfxsize=12cm\epsfbox{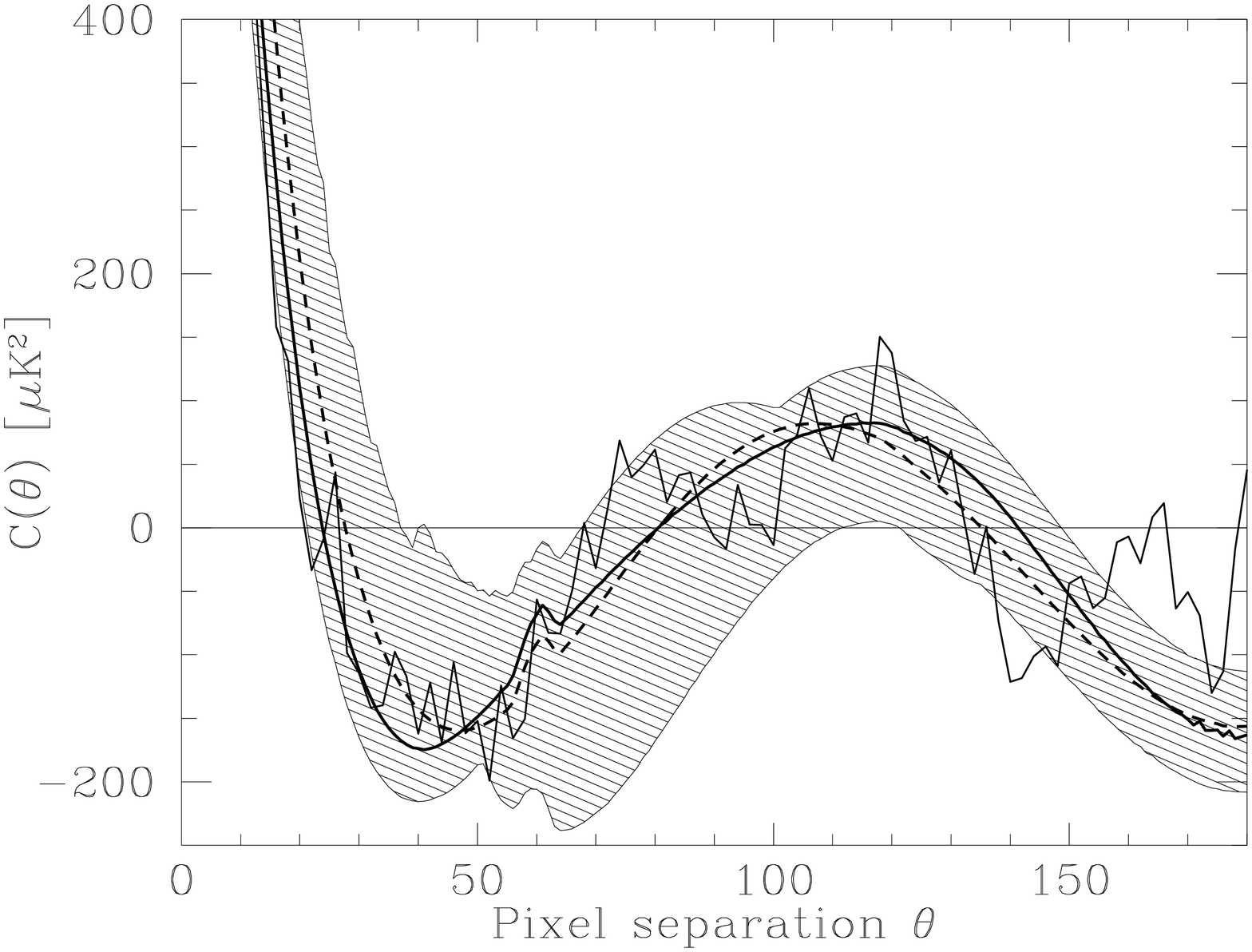}}
\caption{
How dipole removal alters the correlation function.
}
The pixel correlation $\expec{x_i x_j}$ is plotted as a function 
of the angle between the two pixels.
The dashed curve shows the naive correlation function 
corresponding to $n=1$, $\Q=18\mK$.
The shaded region shows the range of correlations actually 
occurring after the monopole, dipole and quadrupole are removed 
outside of a $20^{\circ}$ Galactic cut, the heavy 
solid curve showing the average correlation.
The wiggly line is the correlation function
naively extracted from the 53+90 GHz 2 year COBE data.
The bumps around $\theta=60^{\circ}$ are due to the
noise correlation reported by Lineweaver {\etal} (1994).

\label{CorrFig}
\end{figure}

The standard way to eliminate the influence of the nuisance term 
is to ``remove" the monopole, dipole and perhaps quadrupole
from the data $\vx$ before performing an analysis.
Although often not thought of in such a way, this 
is in fact a linear operation, corresponding to 
multiplying the
data by a certain matrix $D$. 
Let us define the matrix $\Zp$ to be an orthonormalized 
version of $Z$, {\ie}, a matrix whose columns are orthonormal and
span the same space as the columns of $Z$.
This can be achieved either by a standard technique such 
as Gram-Schmidt orthogonalization or singular-value decomposition,
or by simply defining
$\Zp\equiv Z(Z^t Z)^{-1/2}$
and computing the square root by Cholesky decomposition.
In either case, the $N\times N$ matrix $\Zp\Zp^t$, which has rank 
$(\lstar+1)^2$, acts as a projection operator onto the space spanned by the 
multipoles with $\l\leq\lstar$, and it is clear that 
the ``removal matrix" $D$
is given by
$D \equiv I - \Zp\Zp^t$.
Thus defining the {\it corrected data} as
\beq{xpEq}
\vxp \equiv D\vx,
\eeq
it is readily seen that 
$\expec{\vxp} = D Z\vq = 0$,
so that the effect of the nuisance term has been eliminated.
The covariance matrix for the corrected data is simply
\beq{MpEq}
\Mp\equiv \expec{\vxp\vxp^t} = DMD^t.
\eeq
We wish to stress that
the covariance matrix for the corrected data can in general
{\it not} be described by a correlation function
(Bennett {\etal} 1994; Wright {\etal} 1994b). 
After the correction, the correlation 
$M_{ij}$ between two pixels does not depend merely on the 
angle cosine between them, $\cos(\theta) = \nh_i\cdot\nh_j$,
but also on the Galactic latitude of both pixels. 
In other words, the monopole and dipole removal breaks the
rotational symmetry of the correlation function.
This is due to the
well-known fact that the Galactic cut destroys the 
orthogonality of the various multipoles, {\ie}, $Y^t Y\neq I$.
Hence when the monopole is removed, other multipoles are affected
-- primarily the $m=0$ components of the
quadrupole and the hexadecupole.
The dipole removal strongly affects the three components of the
octupole that have $|m|\leq 1$, {\etc}
The situation is illustrated in
\fig{CorrFig}.

Some of the first papers analyzing the COBE data computed
the average correlation between pixels in various bins of 
angular separation. For the 53+90 GHz 2 year data, this
produces the wiggly line in \fig{CorrFig}. Cosmological
parameters were then fitted by comparing this to the 
theoretical, rotationally symmetric
correlation function given by \eq{Meq} (the dashed line). 
As we have seen, this is not quite the correct thing to do. 
The wiggly line should be compared with the theoretical 
prediction for the same quantity (heavy solid line), 
using \eq{MpEq}.
Monte Carlo simulations ({\eg} Seljak \& 
Bertschinger 1993;  Bennett {\etal} 1994)
have shown that this effect leads to a small but non-negligible
bias.

In addition, the figures show that the symmetry breaking is quite
substantial, the correlations for any given angular separation varying 
within the shaded region.
As the correlation function method throws away this information about 
azimuthal dependence
by averaging in bins for fixed $\theta$, 
one might expect the resulting error
bars to be slightly larger than optimal.

For the reader with computational interests, we point out that 
the correction of $M$ is quite a rapid procedure. No 
$N\times N$ matrices need to be multiplied together,
since
\beq{SpeedEq}
\Mp = DMD^t = M - (ZF^t+FZ^t) + Z(Z^tF)Z^t,
\eeq
where $F\equiv MZ$.

\section{Results}

\begin{figure}[phbt]
\centerline{\epsfxsize=16cm\epsfbox{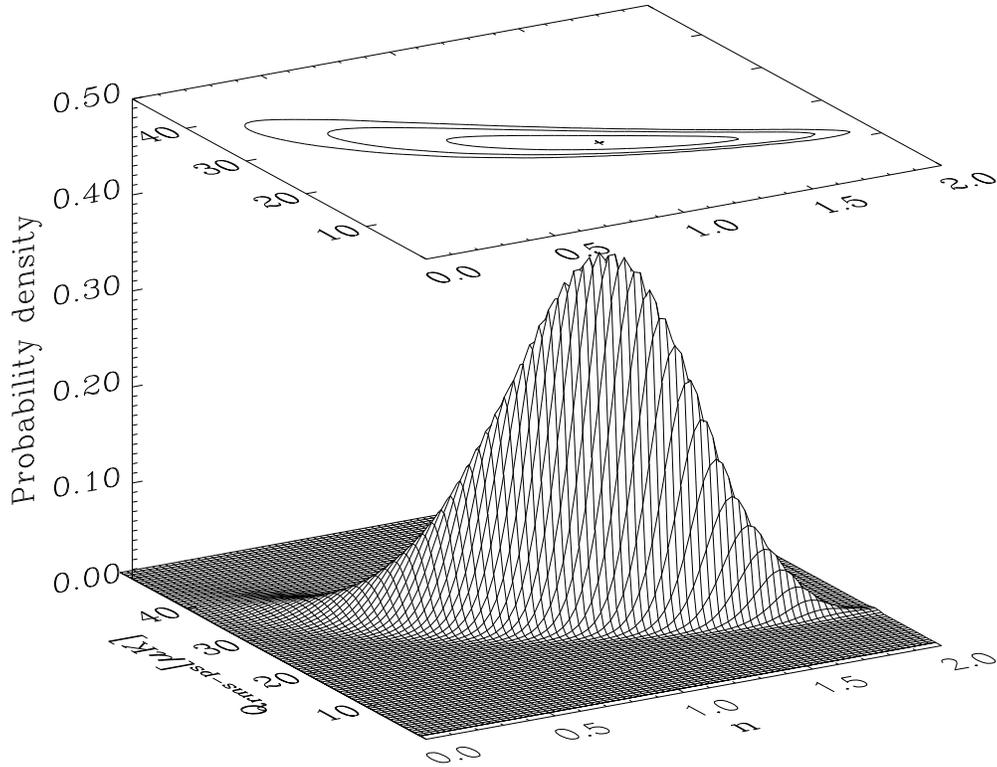}}
\caption{
The normalized likelihood.
}
The Bayesian probability distribution for the spectral index
$n$ and the normalization $\Q$ is plotted (bottom)
using the 
combined 53 and 90 GHz 
two year COBE data and a uniform prior.
The three contours (top) 
show the areas containing $68\%$, $95\%$ and 
$99\%$ of the probability, respectively.
The ``+" shows the maximum-likelihood estimate,
$(n,\Q) \approx (1.15,18.2\mK)$.

\label{3DFig}
\end{figure}

\begin{figure}[phbt]
\centerline{\epsfxsize=12cm\epsfbox{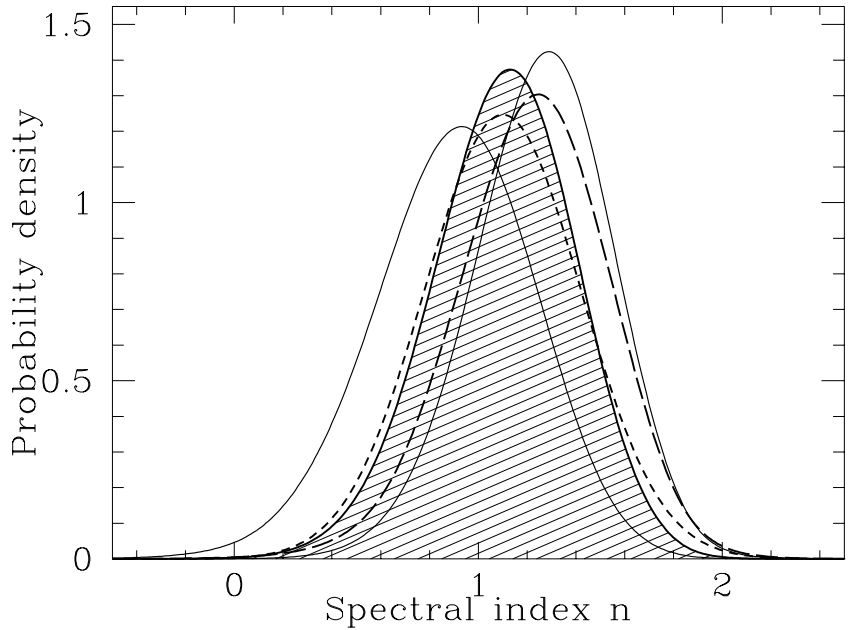}}
\caption{
Marginal likelihoods for $n$.
}
In order of decreasing peak height,
the curves are obtained by the naive brute force
method (not corrected for ``dipole bias"), the brute
force method (shaded), Bunn \& Sugiyama 1994 (dashed), 
G\'orski {\etal} 1994 (dotted) and the brute force method with
quadrupole removed, respectively.
All five curves are based on the combined 53 and 90 GHz 
two year COBE data and are marginalized over the normalization at
the ``pivot point". 
All but the last curve include the quadrupole.

\label{MargFig}
\end{figure}

Assuming that the cosmic signal $\va$ is Gaussian
(that it is a random vector whose probability distribution is 
a multivariate Gaussian), so is $\vxp$.
Thus given an observed data set $\vxp$, the likelihood $\L$
is given by
\beq{Leq}
-2\ln\L = \ln\det\Mp + \vxp^t\Mp^{-1}\vxp,
\eeq
up to an uninteresting additive constant. 
For a spectrum due to Sachs-Wolfe fluctuations
from power law density perturbations,
\beq{ClEq}
C_l =  
\left({4\pi\over 5}\right)
{\Gamma\left({9-n\over 2}\right)
\over
\Gamma\left({3+n\over 2}\right)}
{\Gamma\left(l+{n-1\over 2}\right)\over
\Gamma\left(l+{5-n\over 2}\right)}
\Q^2,
\eeq
and the likelihood 
$\L(n,\Q)$ becomes 
a function of merely two parameters, the spectral index
$n$ and the quadrupole normalization $\Q$.
($\Q$ is denoted $\Qrmsps$ in many papers.)
We have evaluated $\L$ numerically on a grid of points, and 
the resulting normalized likelihood is shown in 
\fig{3DFig}.
With $\lstar=1$ (monopole and dipole removed), 
the maximum likelihood estimate is 
$(n,\Q) = (1.15,18.2\mK)$,
and with $\lstar=2$ (quadrupole removed as well), it is
$(n,\Q) = (0.95,21.3\mK)$.
The normalized marginal likelihood for $n$ is plotted in 
\fig{MargFig}
(the shaded distribution),
together with that obtained in G94 and BS95.
As in those papers, we used the combined 53 and 90 GHz maps
(including also the 31 GHz map gave us almost no
error bar reduction, as its noise level is so much higher),
and a uniform prior.
We use pixel by pixel minimum variance weighting when 
combining the data sets for different channels 
and frequencies. In other words, 
a pair of data sets
$\{x'_i,\sigma'_i\}$ and
$\{x''_i,\sigma''_i\}$
are combined into a new data set 
$\{x_i,\sigma_i\}$ according to the formulas
\beqa{ComboEq}
x_i &=&{{\sigma'}_i^{-2} x'_i + {\sigma''}_i^{-2} x''_i\over
{\sigma'}_i^{-2} + {\sigma''}_i^{-2}},\\
\sigma_i &=& \left({\sigma'}_i^{-2} + {\sigma''}_i^{-2}\right)^{-1/2}.
\eeqa
As expected, our brute force technique gives the smallest error bars,
{\ie}, the narrowest distribution, corresponding to the highest peak.
Also in agreement with our expectations,  the other 
techniques are seen to be fairly close to optimal, 
with only marginally broader distributions.
Our $1\sigma$ confidence intervals are
$n=1.10\pm 0.29$, $\Q=(20.2\pm 4.6)\mK$.
For the former, we have followed G94 
in marginalizing over the 
normalization at the ``pivot point", in our case $C_9$. 
Marginalizing over $\Q$ instead (which simply corresponds
to using a different Bayesian prior) makes only a minor 
difference: $n=1.07\pm 0.30$. Conditioning on $n=1$, 
$\Q = (20.3\pm 1.5)\mK$.
Choosing $\lstar=2$ (removing the quadrupole as well) is seen to
yield a lower $n$-estimate, $n=0.90\pm 0.32$,
just as reported by 
other authors (since the observed quadrupole is 
relatively small, it tends to favor power spectra
with greater slope).
Note that quadrupole removal also increases the error bars, as 
it amounts to throwing out a considerable amount of cosmological
information.

In computing the curves in 
\fig{MargFig},
we have assumed uncorrelated noise.
We also computed the $\lstar=1$ curve using
the 1st year noise correlation given by
Lineweaver {\etal} (1994),
but omit this from the plot 
as the difference is so small that it
is hardly visible. The combined 2 year noise correlation is of 
course weaker still.
In other words, uncorrelated noise is an excellent approximation.

In contrast, the approximation that 
the correlations are not affected by monopole and dipole removal
turns out not to be very good,
which is hardly surprising in view of \fig{CorrFig}.
The shaded curve was computed by removing the monopole and dipole from the data
and using the corrected covariance matrix given by \eq{MpEq}. 
The curve resulting from using the naive covariance matrix $M$ instead
is seen to peak 
further to the right, the approximation causing a shift 
of about the same magnitude as the quadrupole removal, but in
the opposite direction.
We carried out Monte Carlo simulations with 
$(n,\Q) = (1,20\mK)$ fake skies, which verified that 
the latter approximation was biased high whereas
the exact treatment appeared to be fairly unbiased.
%
%
It is easy to understand the sign of this effect on physical grounds:
since the non-orthogonal multipoles couple mainly 
to $l$-values differing by small even numbers, removing 
the monopole and dipole covertly removes parts of other
low multipoles, which increases the slope of the best fit power
spectrum.  

In linear techniques, this effect can be removed simply
by proper treatment of the covariance matrix; in nonlinear techniques
one must resort to Monte Carlo techniques.  All of the 
analyses that quote estimates of $n$ make some such attempt to 
account for potential bias.
Bennett et al. (1994)
have reported and removed just such a bias from
their estimates.  
Smoot et al. (1994) check for bias with Monte Carlo simulations
and find none.  Wright et al. (1994b) use Monte Carlo simulations to
compute the correct covariance matrix for their multipole
estimates, and argue that this is 
sufficient to remove bias.

We also tested an alternative approach to handling the bias 
from dipole subtraction, used in BS95. 
If one uses the uncorrected data $\vx$
instead of $\vxp$, the likelihood function becomes
\beq{Leq2}
-2\ln\L = \ln\det M + (\vx-Z\vq)^t M^{-1}(\vx-Z\vq),
\eeq
where the unknown multipoles $\vq$ remain as nuisance parameters.
We computed the marginal distribution over
$(n,\Q)$ by simply integrating $L$ over all $\vq$.
This integral can readily be done analytically, 
and is of course independent of 
the value of the unknown nuisance multipoles. The result was 
virtually indistinguishable from that of the other method, and is 
therefore
not plotted.

For readers with computational interests, we
conclude this section with a few practical comments.
The $N\times N$ matrix $M$ is quite well-conditioned, due to
the large noise contributions to its diagonal,
and can be readily Cholesky decomposed. With an optimized code, 
only $N(N+1)/2$ numbers need to be stored in RAM, as it is symmetric
and the result can be computed in such an order as to 
gradually overwrite the original matrix.
On a good (1994) workstation, this takes about ten minutes with $N=4038$.
The matrix $\Mp$, on the other hand, 
is singular, as it by construction has rank
$N-(\lstar+1)^2$.  (The pixels are not independent
when we know that their mean is zero, their dipole
is zero, {\etc})
Thus the statistically
correct thing to do is to throw away $(\lstar+1)^2$ arbitrary
pixels
before the likelihood analysis.
The resulting matrix inversion is not as well-conditioned as that 
for $M$, so whereas single precision suffices for $M$, double 
precision should be used here. 
As a stability test, we repeated the analysis with an additional 
100 random pixels thrown away, which greatly increases the condition
number of the covariance matrix, and obtained virtually identical results.

\section{Discussion}

In this letter, we have carried out a ``brute force" analysis of 
the two-year COBE
DMR data which is optimal in the sense of giving the smallest
possible error bars.
The results, $n=1.10\pm 0.29$, agree well with previous work, reinforcing
the conclusion that the large-scale CMB fluctuations are
consistent with the standard inflationary scenario.

\noindent
We draw the following conclusions:

\begin{enumerate}

\item
All the published linear methods of estimating the COBE power spectrum
(B95, G94, BS95)
are close to optimal, in the sense 
of giving error bars near the theoretical lower limit.

\item 
The routine removal of the monopole and dipole 
can introduce a bias, tending to lead to a slight 
overestimate of the spectral index $n$.
In linear techniques, this problem can be simply dealt with by using
the appropriate covariance matrix. When using the
(quadratic) correlation function technique, either the theoretical 
correlation function should be corrected as shown in
\fig{CorrFig}, or Monte-Carlo simulation should be used 
to subtract the resulting bias. Published work using the
latter approach agrees well with our results.

\item
It is well-known that
the slight noise correlation 
reported by Lineweaver {\etal} (1994) 
has only a small impact on CMB results.
We have evaluated this impact numerically, by including the 
full noise-covariance matrix, and 
confirmed that the effect is negligible even at the high accuracy
used here.

\item
The quadratic methods have the advantage of
giving sharp constraints with very few $(\simlt 100)$
reduced data points, which also tend to be easy to interpret
physically (correlation function, multipoles). 
One drawback is that their probability
distribution is no longer Gaussian, and if a
Gaussian approximation is made, the 
computation of their covariance matrix 
can be rather cumbersome (Seljak \& Bertschinger 1993;
Wright {\etal} 1994b; de Oliveira-Costa \& Smoot 1995) compared to the
linear case.

\item 
The above techniques all assume Gaussian fluctuations.
Although topological methods may not be
the most efficient way to constrain the power
spectrum, they will no doubt provide very interesting
tests of the Gaussianity hypothesis as angular
resolution improves.

\end{enumerate}
In the future, as much larger data sets may become 
available through proposed experiments such as the COBRAS/SAMBA
satellite, the brute-force approach used here will hardly be feasible.
It is thus encouraging that the faster methods reviewed
give results that are so close to optimal.

\bigskip
The authors wish to thank Kryzstof G\'orski, Dag Jonsson,
Douglas Scott,
George Smoot, and Philip Stark
for useful comments, Charley Lineweaver for providing
the numerical noise correlation data, and 
Ang\'elica de Oliveira Costa for help with IDL plots.
The COBE data sets were developed by the NASA
Goddard Space Flight Center under the guidance of the COBE Science
Working Group and were provided by the NSSDC.



\section{References}

\rf Banday, A. J. {\etal} 1994;ApJ;436;L99

\rf Bennett, C.L. {\etal} 1992;ApJ;396;L7

\rf Bennett, C.L. {\etal} 1994;ApJ;436;423

\rg Bond, J. R. 1995;Phys. Rev. Lett.;74;4369;B95
     
\rf Bunn, E.F., Scott, D., \& White, M. 1995;ApJ;441;L9

\rg Bunn, E.F. \& Sugiyama, N. 1995;ApJ;446;49;BS95

\rf de Oliveira Costa, A. \& Smoot, G.F. 1995;ApJ;448;477 
 
\rf Fisher, R. A. 1935; J. Roy. Stat. Soc.;98;39

\rf G\'orski, K. M. 1994;ApJ;430;L85

\rn G\'orski, K. M. {\etal} 1994, {\it ApJ}, {\bf 430}, L89 (``G94").

\rf Gould, A. 1993;ApJ;403;L51

\rf Hinshaw, G. {\etal} 1994;ApJ;431;1

\rn Karhunen, K., {\it \"Uber lineare Methoden in der 
Wahrscheinlichkeitsrechnung} (Kirjapaino oy. sana, Helsinki, 1947).

\rn Kenney, J. F. \& Keeping, E. S. 1951, {\it Mathematics of Statistics, Part II},
2nd ed. (Van Nostrand, New York).

\rf Kogut, A. {\etal} 1995;ApJ;439;L29

\rf Lineweaver, C.H., Smoot, G.F., Bennett, C.L., Wright, E.L., Tenorio, L.,
Kogut, A., Keegstra, P.B., Hinshaw, G., \& Banday, A.J. 1994;ApJ;436;452


\rf Seljak, U. \& Bertschinger, E. 1993;ApJ;417;L9


\rf Smoot, G. F. {\etal} 1992;ApJ;396;L1

\rf Smoot, G. F. {\etal} 1994;ApJ;437;1

\rf Torres, S. 1994;ApJ;423;L9


\rf Scaramella, R. \& Vittorio, N. 1993;MNRAS;263;L17


\rf White, M. \& Bunn, E.F. 1995; ApJ;450;477
     
\rf Wright, E. L. {\etal} 1994a;ApJ;420;1

\rf Wright, E. L. {\etal} 1994b;ApJ;436;443


\setcounter{secnumdepth}{-1}
\section{Appendix}

In this Appendix we prove the statement that linear data compression
can never give a likelihood function that is more sharply peaked, on
average, than the likelihood function of the uncompressed data.

Let us begin by establishing some notation.  Let $\vx$ be an
$N$-dimensional vector representing the uncompressed data.  We assume
as usual that $\vx$ is Gaussian distributed with zero mean, and we
denote its covariance matrix $\langle \vx \vx^T\rangle$ by $M_x$.  Let
$A$ be an $N'\times N$ ``compression matrix'' (with $N'<N$), and let
$\vy=A\vx$ be the $N'$-dimensional compressed data vector.

Since $\vx$ is a Gaussian random vector, its likelihood function
is given by
\beq{gausslike}
\Lambda_x=\vx^TM_x^{-1}\vx+\ln\det M_x=\mbox{Tr}\left(M_x^{-1}\vx\vx^T
+\ln M_x\right),
\eeq
where $\Lambda_x=-2\ln L_x$.  (Throughout this Appendix, we will
find it convenient to denote explicitly by subscripts the random variable
with respect to which we are computing likelihoods.)
Suppose we are trying to estimate some
parameter $q$.  We wish to compute the ``width'' of the likelihood
function,
viewed as a function of $q$.  We quantify this width as follows.
Compute $L_x$ as a function of $q$ and find its maximum.  Near the
maximum, we can approximate $\ln L_x$ as a quadratic, and we see that
the width of the peak is inversely 
proportional to the square root of the second derivative of $\ln
L_x$ evaluated at the peak.  Our concern will be with the average width
of the likelihood function, so we will be interested in the parameter
\beq{gammadef}
\gamma_x=\langle \Lambda_x''\rangle,
\eeq
where primes denote derivatives with respect to $q$ and the derivative
is evaluated at the point where $\Lambda_x'=0$.  (This is the same
parameter that was adopted for the optimization problem in BS95.)
We want to show that
replacing the full data vector $\vx$ by the compressed data
$\vy$ can never cause $\gamma$ to decrease.  In other words, we 
want to prove that
\beq{gammaineq}
\gamma_y\ge\gamma_x.
\eeq

We begin by extending the matrix $A$ to a nonsingular $N\times N$ matrix
$\tilde A$ as follows.  Add $N-N'$ rows to the bottom of $A$, making
sure that the added rows are linearly independent of each other
and of the rows of $A$, and are orthogonal to the rows of $A$ with
respect to the inner product $\langle {\bf p},{\bf q} \rangle\equiv
{\bf p}M_x{\bf q}$.\footnote{
This is always possible.  Since $M_x$ is positive definite, the function
$\langle\cdot,\cdot\rangle$ is a perfectly good inner product, and
we can therefore perform ordinary Gram-Schmidt orthogonalization to
generate the extra rows.}
Let $\tilde\vy\equiv\tilde A\vx$.  $\tilde\vy$ is an $N$-dimensional vector
whose first $N'$ elements are the elements of $\vy$.  Let $\vz$ be a
vector consisting of the other $N-N'$ elements of $\tilde\vy$:
\beq{xxxxx}
\tilde\vy=\left(\matrix{\vy \cr
 \vz}\right).
\eeq

Since $\vx$ and $\tilde\vy$ are related via a nonsingular linear 
transformation, the likelihoods derived from them are the same,
up to an overall multiplicative constant.\footnote{
This is easy to see by direct computation: the covariance matrix of $\vy$
is $M_y=AM_xA^T$, so
$\Lambda_y=\vy^T M_y^{-1}\vy+\ln\det M_y=\vx^T A^T(AM_xA^T)^{-1} A\vx
+\ln\det(AM_xA^T)=\vx^TM_x^{-1}\vx+\ln\det M_x+2\ln\det A
=\Lambda_x+2\ln\det A$.}
Furthermore, because of the orthogonality condition we have imposed,
the covariance matrix of $\tilde\vy$ is block diagonal:
\beq{x}
M_{\tilde y}=\left(\matrix{M_y & 0 \cr
0 & M_z}\right).
\eeq
The likelihood $L_{\tilde y}$ therefore factors:
\beq{asdvgsadf}
L_{\tilde y}=L_yL_z
\eeq
Since $\gamma$ is linear in $\ln L$, we know that $\gamma_{\tilde y}=
\gamma_y+\gamma_z$.  Since $\gamma_x=\gamma_{\tilde y}$, all we have
to do to prove the inequality (\ref{gammaineq}) 
is show that $\gamma_z\ge 0$.
This follows immediately from the following relation, proved in BS95:
\beq{xxx}
\gamma=\mbox{Tr}\left((M^{-1}M')^2\right).
\eeq
Since $\gamma_z$ is the trace of the square of a matrix, it can never
be negative.


\def\fheight{10.3cm} \def\fwidth{14.5cm}

\newpage

\newpage

\newpage

\end{document}